\documentclass[aps,twocolumn]{revtex4}
\usepackage{graphicx}
\usepackage{epsfig}
\usepackage{color}
\begin{document}

\title{Pure spin photocurrents}
\author{E L Ivchenko and S A Tarasenko}
\affiliation{A F Ioffe Physico-Technical Institute, Russian
Academy of Sciences, 194021 St Petersburg, Russia}

\begin{abstract}
The pure spin currents, i.e., the counterflow of particles with
opposite spin orientations, can be optically injected in
semiconductors.  Here, we develop a phenomenological theory, which
describes the polarization dependencies of spin currents excited
by linearly polarized light in bulk semiconductors and quantum
well structures of various symmetries. We present microscopic
descriptions of the pure spin photocurrents for interband optical
transitions in undoped quantum wells as well as for direct
intersubband and indirect intrasubband (Drude-like) transitions in
$n$-doped quantum well structures. We also demonstrate that pure
spin currents can be generated in structures of sufficiently low
symmetries by simple electron gas heating. The theoretical results
are compared with recent experimental observations.
\end{abstract}
\maketitle

\section{Introduction}\label{Introduction}

\vspace{-0.2cm}

By definition, the pure spin current of free carriers, electrons
or holes, is a spin flux without an electric current. It can be
conceived as formed by opposing equivalent flows of spin-up and
spin-down particles. This non-equilibrium distribution of carriers
in the wave vector and spin spaces is characterized by zero charge
current, because electric currents contributed by spin-up and
spin-down particles cancel each other, but leads to the
accumulation of opposite spins at the opposite edges of the
sample. In particular, a pure spin current is produced in a
direction perpendicular to an applied electric field due to
spin-dependent skew or side-jump scattering of electrons by
impurities or phonons. This effect known as the spin Hall effect
was predicted in early theoretical
studies~\cite{Dyakonov71,Hirsch99} and observed in recent years,
see~\cite{Awschalom04,Wunderlich05,Tinkham06}.

In this study, we present the theory of pure spin currents
generated under the light absorption in an unbiased semiconductor
structures. In spin-dependent optical spectroscopy, the efforts
were mostly directed towards (i) the photogeneration of
nonequilibrium spin polarization of carriers, the effect known as
the optical orientation (see, e.g., \cite{optor,ivchbook}), and
(ii) the generation of spin-sensitive electric currents known as
the circular photogalvanic, spin-galvanic and magneto-gyrotropic
effects~\cite{ivchbook,Belkov05}. In contrast, the free carriers
participating in a pure spin current neither have a net spin
polarization nor produce a net charge current. A spacial
separation of electron spins caused by the spin photocurrent was
first observed by using two-color optical coherence control
techniques, due to quantum interference of one- and two-photon
absorption of two orthogonally-polarized overlapping laser pulses
with frequencies $\omega$ and
$2\omega$~\cite{Stevens03,Huebner03,Driel06}. Then, Bhat et
al.~\cite{Bhat05} and Tarasenko and Ivchenko~\cite{TarIvch05}
showed that merely one-photon absorption of linearly polarized
light should produce pure spin currents  in noncentrosymmetric
bulk semiconductors and quantum well (QW) structures: in this case
there is no net motion of charge but spin-up and spin-down
photoelectrons travel in the opposite directions. The theoretical
prediction was followed by an observation of pure spin currents
induced by a single linearly polarized optical pulse in
(110)-oriented GaAs QWs~\cite{Zhao05}.

Theoretically, the two-color generation and control of spin
currents have been extensively
analyzed~\cite{Bhat05,Bhat00,Marti04,Duc05}. Here we will
consider, in order, the one-photon generation of pure spin
currents in unbiased structures under interband, intersubband and
intraband absorption of linearly polarized or unpolarized light
and derive equations for the corresponding currents. Particularly,
we compare different mechanisms of pure spin currents and show
difference in the behavior of their contributions as a function of
the light frequency and the polarization direction with regard to
the crystallographic axes. We also show that pure spin currents
emerge in QW structures as soon as the electron gas is simply
driven out of thermal equilibrium with the crystal lattice.
Finally, we will discuss a new phenomenon which can be called the
pure valley-orbit current and observed in many-valley
semiconductors. The role of spin-up and spin-down states in pure
spin currents is replaced in the valley-orbit current by the index
of the conduction-band valleys: the valleys are equally populated,
there is no net charge current, but the electrons in different
valleys travel in different directions.

\section{Phenomenology}\label{Phenomenology}

Phenomenologically, the spin flux, or, in general, the flux of
angular momentum, is described by a second-rank pseudotensor
$\mathbf{J}$ whose components $J_{\beta}^{\alpha}$ stand for the
flow in the $\beta$ direction of spins oriented along $\alpha$,
with $\alpha$ and $\beta$ being the Cartesian coordinates. Nonzero
components of the photo-induced spin current $\mathbf{J}$ are
determined by the light polarization and the explicit form of
spin-orbit interaction governed by the structure symmetry. They
can be revealed from the symmetry analysis which requires no
knowledge about microscopical mechanisms of the spin current
generation. Indeed, in the regime of linear dependence of
$\mathbf{J}$ on the light intensity $I$, the spin-photocurrent
components are phenomenologically related by
\begin{equation}\label{J_phen}
J_{\beta}^{\alpha} = I \sum_{\gamma\delta}
Q_{\alpha\beta\gamma\delta} \, e_{\gamma} e_{\delta}^* \:,
\end{equation}
to the light-polarization unit vector $\mathbf{e}$ and the complex
conjugate vector $\mathbf{e}^*$. Equation~(\ref{J_phen})
represents the most general form of the spin photocurrent
description because the set of quadratic terms $e_{\gamma}
e_{\delta}^{*}$ fully determines the light polarization
state~\cite{sc_drag}.

Equation~(\ref{J_phen}) can be usefully rewritten in the
equivalent form
\begin{equation}\label{J_phen2}
J_{\beta}^{\alpha} = I \sum_{\gamma\delta}
L_{\alpha\beta\gamma\delta} \frac{e_{\gamma}e_{\delta}^* +
e_{\delta}e_{\gamma}^{*}}{2} + I \sum_{\gamma} C_{\alpha\beta\mu}
{\rm i} [\mathbf{e} \times \mathbf{e}^*]_{\mu} \:,
\end{equation}
where $L_{\alpha\beta\gamma\delta}=(Q_{\alpha\beta\gamma\delta}+
Q_{\alpha\beta\delta\gamma})/2$ is a fourth-rank pseudotensor
symmetric in the two last indices,
$C_{\alpha\beta\mu}=\sum_{\gamma\delta}Q_{\alpha\beta\gamma\delta}
\,\epsilon_{\gamma\delta\mu}/(2 {\rm i})$ is a third-rank tensor,
and $\epsilon_{\gamma\delta\mu}$ is the completely antisymmetric
third-rank pseudotensor, or the Levi-Civita tensor. The
pseudotensor $\mathbf{L}$ describes spin photocurrents which are
independent of the sign of light circular polarization for
elliptically polarized light and can be conveniently measured for
the linearly polarized radiation. In contrast, the tensor
$\mathbf{C}$ stands for helicity-sensitive spin photocurrents
which reverse their polarity upon switching the sign of circular
polarization. This occurs because the cross product ${\rm i}
[\mathbf{e} \times \mathbf{e}^*]$ is zero for linearly polarized
light and proportional to the light helicity for elliptical or
circular polarization. Usually, the absorption of circularly
polarized light results in a considerable spin polarization of
photoexcited carriers~\cite{optor} masking the observation of pure
spin currents. Therefore, in what follows, we focus on spin
currents excited by linearly polarized light only and assume the
polarization vector $\mathbf{e}$ to be real.

In crystals having a zinc-blende structure and characterized by
the symmetry point group $T_d$, the linearly polarized light can
induce both diagonal $J_{\alpha}^{\alpha}$ and off-diagonal
$J_{\beta}^{\alpha}$ ($\alpha \neq \beta$) components of the spin
current. Their polarization properties are described by
\begin{eqnarray}\label{J_phenTd}
J_{\alpha}^{\alpha} = L_1 \, I (e_{\alpha+1}^2 - e_{\alpha+2}^2) \:, \\
J_{\alpha + 1}^{\alpha} = - J_{\alpha}^{\alpha + 1} = L_2 I \,
e_{\alpha}e_{\alpha+1} \:. \nonumber
\end{eqnarray}
Here, $I$ is the light intensity, the index $\alpha$ runs over the
cubic axes $x \| [100]$, $y \| [010]$, and $z \| [001]$, and the
index $\alpha + 1$ is obtained by the cyclic permutation of the
indices $x, y, z$. Note that nonzero values of the
phenomenological parameters $L_1$ and $L_2$ in
Eq.~(\ref{J_phenTd}) are allowed in noncentrosymmetric crystals of
the T$_d$ symmetry and forbidden for diamond-type centrosymmetric
crystals.

The symmetry of (001)-oriented QWs grown from zinc-blende-type
semiconductors reduce to the point group D$_{2d}$ in symmetrical
structures and $C_{2v}$ in asymmetrical structures. For the
latter, the spin photocurrent components photoinduced in the
$(xy)$ plane are described by 10 linearly independent constants as
follows
\begin{eqnarray}\label{J_phenC2v}
&& J_x^x/I = L_1^{\mathrm{B}} e_x^2 + L_2^{\mathrm{B}} e_y^2 +
L_3^{\mathrm{B}} e_z^2 + L_1^{\mathrm{S}} e_x e_y \:, \\
&& J_y^x/I = L_2^{\mathrm{S}} e_x^2 + L_3^{\mathrm{S}} e_y^2 +
L_4^{\mathrm{S}} e_z^2 + L_4^{\mathrm{B}} e_x e_y \:, \nonumber \\
&& J_x^y/I = -L_3^{\mathrm{S}} e_x^2 - L_2^{\mathrm{S}} e_y^2 -
L_4^{\mathrm{S}} e_z^2 - L_4^{\mathrm{B}} e_x e_y \:, \nonumber \\
&& J_y^y/I = -L_2^{\mathrm{B}} e_x^2 - L_1^{\mathrm{B}} e_y^2 -
L_3^{\mathrm{B}} e_z^2 - L_1^{\mathrm{S}} e_x e_y \:, \nonumber \\
&& J_x^z/I = L_5^{\mathrm{B}} e_x e_z + L_5^{\mathrm{S}} e_y e_z
\:, \nonumber \\
&& J_y^z/I = - L_5^{\mathrm{S}} e_x e_z - L_5^{\mathrm{B}} e_y e_z
\:. \nonumber
\end{eqnarray}
Here, the superscript $\mathrm{B}$ marks those coefficients which
are allowed in QWs of the D$_{2d}$ symmetry and can be related to
bulk inversion asymmetry (BIA) of the host crystal and/or
anisotropy of the chemical bonds at the QW interfaces, while the
superscript $\mathrm{S}$ marks the contributions which appear
because of structure inversion asymmetry (SIA) only. Therefore, in
symmetrical (001)-grown QWs, the coefficients $L_{i}^{\mathrm{S}}$
vanish and the polarization dependencies of spin current
components are completely determined by the terms proportional to
$L_{i}^{\mathrm{B}}$. In the opposite limit, where the SIA
predominates and the QW structure can effectively be described by
the axial point group $C_{\infty v}$, the spin photocurrent is
given by Eq.~(\ref{J_phenC2v}) with the BIA-related terms being
disregarded and the coefficients $L_{i}^{\mathrm{S}}$ satisfying
the relation $L_{3}^{\mathrm{S}} = L_{1}^{\mathrm{S}} +
L_{2}^{\mathrm{S}}$.

It follows from Eq.~(\ref{J_phenC2v}) that, under normal incidence
on a (001)-grown QW, the linearly polarized light can excite
fluxes of electron spins oriented only in the interface plane. To
create the $J_x^z$ and $J_y^z$ spin current components, which can
cause the spacial profile of the spin density $S_z$, one has to
irradiate the QW in the oblique-incidence geometry. This is in
contrast to QWs grown along low-symmetry crystallographic axes,
where the normally-incident light can induce fluxes of both the
in-plane and out-of-plane components of the spin polarization.

As an example of such low-symmetry structures, we consider QWs
grown on (110)-oriented substrates and use the $(x', y', z')$
coordinate frame with $z'$ along the growth direction and the
in-plane axes $x' \| [1\bar{1}0]$ and $y' \| [00\bar{1}]$.
Asymmetrically-grown (110)-oriented QWs have the point group $C_s$
and contain only two symmetry elements: the identity and a mirror
plane $m_1 = (1\bar{1}0)$ perpendicular to the $x'$ axis. In this
particular case, components of the spin current excited by
normally-incident light are phenomenologically given by
\begin{eqnarray}\label{J_phenCs}
J_{x'}^{x'}/I = L'_1 e_{x'}e_{y'} \:,\;\; J_{y'}^{x'}/I = L'_2
+ L'_3 (e_{x'}^2-e_{y'}^2) \:, \\
J_{x'}^{y'}/I = L'_4 + L'_5 (e_{x'}^2-e_{y'}^2) \:,\;\;
J_{y'}^{y'}/I = L'_6 e_{x'}e_{y'} \:, \nonumber \\
J_{x'}^{z'}/I = L'_7 + L'_8 (e_{x'}^2-e_{y'}^2) \:,\;\;
J_{y'}^{z'}/I = L'_9 e_{x'}e_{y'} \:. \nonumber
\end{eqnarray}
Symmetrical (110)-grown QWs contain an additional mirror plane
$m_2 = (110)$ perpendicular to the $z'$ axis. Reflection by the
plane $m_2$ changes the sign of the $J_{\beta}^{x'}$ and
$J_{\beta}^{y'}$ ($\beta=x',y'$) components of the spin current
but does not modify $J_{\beta}^{z'}$ as well as the in-plane
components of the polarization vector $\mathbf{e}$. Therefore, in
symmetrical (110)-grown QWs, the parameters $L'_1 ... L'_6$ vanish
and the spin photocurrent is solely described by the last line of
Eq.~(\ref{J_phenCs}).

Microscopically, the emergence of a pure spin current under the
light absorption is related to spin-orbit interaction coupling
spin states and spatial motion of charge carriers, the latter
being directly affected by the electric field of the light. In
terms of the kinetic theory, the $J_{\beta}^{\alpha}$ component of
the spin photocurrent in the conduction band is contributed by a
non-equilibrium correction $\propto \sigma_{\alpha} k_{\beta}$ to
the electron spin density matrix, where $\sigma_{\alpha}$ is the
Pauli matrix and $\mathbf{k}$ is the wave vector. In general, the
concept of spin currents is uncertain in systems with spin-orbit
interaction, since the spin and spin-dependent velocity cannot be
determined simultaneously (see, e.g., Ref.~\cite{Rashba04}).
Mathematically it is caused by the fact that the Pauli matrices
and the velocity operator do not commute. However, this problem of
the spin current definition emerges in high orders in the
spin-orbit interaction only and vanishes for the cases, where spin
currents are directly proportional to the constant of spin-orbit
coupling. To the first order in the spin-orbit coupling and within
the relaxation time approximation, components of the pure spin
current photoinduced in the conduction band are given by
\begin{equation}\label{Spinflux}
J_{\beta}^{\alpha}=\sum_{\mathbf{k}} \tau_e \mathrm{Tr} \left[
\frac{\sigma_{\alpha}}{2} \, v_{\beta}(\mathbf{k})
\,G(\mathbf{k})\right] \:,
\end{equation}
with the spin-dependent corrections being taken into account
either in the velocity operator $\mathbf{v}(\mathbf{k})$ or in the
photogeneration rate of the spin density matrix $G(\mathbf{k})$.
Here, $\tau_e$ is the relaxation time of the spin current which
can differ from the conventional momentum relaxation time that
governs the electron mobility. Electron-electron collisions
between particles of opposite spins, which do not affect the
mobility, contribute to the relaxation of pure spin currents
reducing the time $\tau_e$ (see, the spin Coulomb
drag~\cite{Amico00,Weber05} and the effect of electron-electron
interaction on spin relaxation \cite{Harley07}, and references
therein).

\section{Interband transitions in QWs}\label{Interband}

Among microscopic mechanisms of the pure spin photocurrent we
first discuss that related to $\mathbf{k}$-linear spin-orbit
splitting of quantum subbands~\cite{TarIvch05}, in the following
the split-subband-related mechanism. The mechanism is most easily
conceivable for direct transitions between the heavy-hole valence
subband $hh1$ and conduction subband $e1$ in (110)-grown QWs. In
such structures, the spin component along the QW normal $z'$ is
coupled with the in-plane electron wave vector due to the terms
proportional to $\sigma_{z'}k_{x'}$ and $J_{z'} k_{x'}$ in the
conduction and valence bands, respectively, where $J_{z'}$ is the
4$\times$4 matrix of the angular momentum $3/2$~\cite{ivchbook}.
This leads to $\mathbf{k}$-linear spin splitting of both the
electron subband $e1$ and the valence subband $hh1$ into branches
with the spin projection $\pm 1/2$ and $\pm 3/2$, respectively, as
sketched in Fig.~1(a). The corresponding dispersions in the
subbands at small in-plane wave vector are given by
\begin{equation}\label{E110}
E_{\mathbf{k},\pm1/2}^{(e1)} = \frac{\hbar^2
(k_{x'}^2+k_{y'}^2)}{2m_e} \pm \gamma_{z'x'}^{(e1)} k_{x'} \:,
\end{equation}
\[
E_{\mathbf{k},\pm 3/2}^{(hh1)} = - \frac{\hbar^2
(k_{x'}^2+k_{y'}^2)}{2m_h} \pm \gamma_{z'x'}^{(hh1)} k_{x'} \:,
\]
where $m_e$ and $m_h$ are the electron and hole effective masses
in the QW plane. Note that the spin splitting of the conduction
subband is relativistic and, therefore, small as compared to the
nonrelativistic term $J_{z'} k_{x'}$ describing the splitting of
heavy-hole states in $(110)$-grown structures.

Due to the selection rules, the allowed direct optical transitions
from the valence subband $hh1$ to the conduction subband $e1$ are
$|+3/2 \rangle \rightarrow |+1/2 \rangle$ and $|-3/2 \rangle
\rightarrow |-1/2 \rangle$~\cite{optor}, as illustrated in
Fig.~1(a) by vertical lines. Under excitation with linearly
polarized or unpolarized light the rates of both transitions are
equal. In the presence of spin splitting, the optical transitions
induced by photons of the fixed energy $\hbar\omega$ occur in the
opposite points of the $\mathbf{k}$ space for the spin branches
$s_{z'}=\pm 1/2$. Such an asymmetry of photoexcitation results in
a flow of electrons within each spin branch. The corresponding
fluxes ${\bf i}_{+ 1/2}$ and ${\bf i}_{- 1/2}$ are of equal
strengths but of opposite directions. Thus, this non-equilibrium
electron distribution is characterized by the nonzero spin current
$(1/2) ({\bf i}_{+1/2} - {\bf i}_{- 1/2})$ but a vanishing charge
current, $e ({\bf i}_{+1/2} + {\bf i}_{- 1/2})=0$.

\begin{figure}[t]
\leavevmode \epsfxsize=0.95\linewidth
\centering{\epsfbox{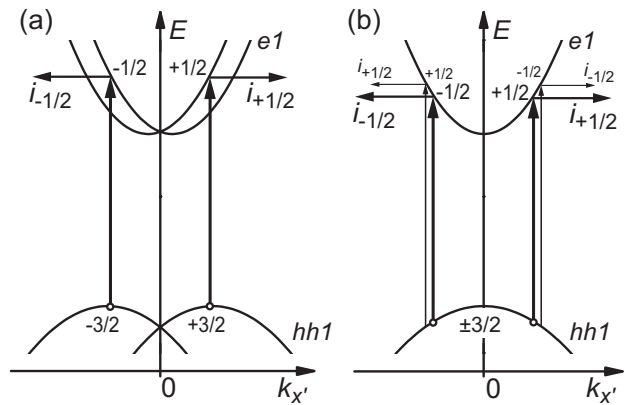}} \caption{Microscopic mechanisms
of the pure spin photocurrent induced by interband excitation with
linearly polarized light in (110)-QWs due to (a)
$\mathbf{k}$-linear spin splitting of subbands and (b)
$\mathbf{k}$-linear terms in the transition rates.}
\end{figure}

To calculate the spin current, we note that the points of optical
transitions in the $\mathbf{k}$ space are determined by the energy
and quasi-momentum conservation which reads
\begin{equation}\label{conservat}
E_g^{QW} + \frac{\hbar^2 (k_{x'}^2+k_{y'}^2)}{2\mu} + 2 s_{z'}
(\gamma_{z'x'}^{(e1)} - \gamma_{z'x'}^{(hh1)}) k_{x'} =
\hbar\omega \:,
\end{equation}
where $E_g^{QW}$ is the QW band gap at $\mathbf{k}=0$ and $\mu=m_e
m_h/(m_e+m_h)$ is the reduced effective mass. Owing to spin
splitting of both the $e1$ and $hh1$ subbands, electrons are
photoexcited into the spin branches $s_{z'}=\pm 1/2$ with the
average velocities
\begin{equation}\label{vx110}
\langle v_{x'} \rangle = \frac{\hbar}{m_e} \langle k_{x'} \rangle
+ 2 s_{z'} \frac{\gamma_{z'x'}^{(e1)}}{\hbar} = 2 s_{z'}
\frac{\mu}{\hbar} \left(\frac{\gamma_{z'x'}^{(e1)}}{m_h}
+\frac{\gamma_{z'x'}^{(hh1)}}{m_e}\right) \:.
\end{equation}
The opposite motion of spin-up and spin-down electrons decays
within the relaxation time $\tau_e$. However, under the
steady-state excitation the electron generation is continuous
resulting in the spin current
\begin{equation}\label{Jcv110a}
J_{x'}^{z'} = \frac{\mu \tau_e}{2 \hbar}
\left(\frac{\gamma_{z'x'}^{(e1)}}{m_h}
+\frac{\gamma_{z'x'}^{(hh1)}}{m_e}\right)
\frac{\eta_{cv}}{\hbar\omega} I \:,
\end{equation}
where $\eta_{cv}$ is the QW absorbance.

Another contribution to the spin photocurrent may come from ${\bf
k}$-linear terms in the matrix elements of the interband optical
transitions~\cite{TarIvch07}, hereafter referred to as the
matrix-element-related mechanism. Taking into account
$\mathbf{k}$$\cdot$$\mathbf{p}$ admixture of the remote conduction
band $\Gamma^c_{15}$ to the valence-band and conduction-band
states $X_{{\bf k}},Y_{{\bf k}},Z_{{\bf k}}$, and $S_{{\bf k}}$,
respectively, one derives the interband matrix elements of the
velocity operator for bulk zinc-blende-type
semiconductors~\cite{Geller89,Khurgin06}
\begin{eqnarray}\label{ev_k}
i\langle S_{{\bf k}} | {\bf e}\cdot{\bf v}| X_{{\bf k}} \rangle =
(P/\hbar) [e_x + i\beta(e_y k_z + e_z k_y)] \:,\\
i\langle S_{{\bf k}} | {\bf e}\cdot{\bf v}| Y_{{\bf k}} \rangle =
(P/\hbar) [e_y + i\beta(e_x k_z + e_z k_x)] \:,\nonumber \\
i\langle S_{{\bf k}} | {\bf e}\cdot{\bf v}| Z_{{\bf k}} \rangle =
(P/\hbar) [e_z + i\beta(e_x k_y + e_y k_x)] \nonumber \:,
\end{eqnarray}
where $P=i(\hbar/m_0)\langle S|p_z|Z\rangle$,
$P'=i(\hbar/m_0)\langle S|p_z|Z'\rangle$ and
$Q=i(\hbar/m_0)\langle X'|p_y|Z\rangle$ are the interband matrix
elements at the $\Gamma$ point of the Brillouin zone, $X', Y', Z'$
are the Bloch functions of the $\Gamma^c_{15}$ band, $m_0$ is the
free electron mass, $\beta = Q P'(2E'_g+E_g)/[PE'_g(E'_g+E_g)]$ is
a material parameter, $E_g$ is the fundamental band gap, and
$E'_g$ is the energy separation between conduction bands
$\Gamma^c_{15}$ and $\Gamma_6$ at the $\Gamma$ point. For GaAs,
the coefficient $\beta$ can be estimated as $0.2\div 1$~\AA\,
depending on the band parameters used~\cite{Jancu05,Pfeffer96}.

The $\mathbf{k}$-linear terms in Eq.~(\ref{ev_k}) do not modify
the selection rules for optical transitions from the heavy-hole
valence subband to the conduction band. As before, the only
allowed transitions are $|+3/2 \rangle \rightarrow |+1/2 \rangle$
and $|-3/2 \rangle \rightarrow |-1/2 \rangle$. However, the rates
of the above transitions become dependent of the in-plane wave
vector. Particularly, for the linearly polarized light normally
incident upon a (110)-grown QW, the squared moduli of the matrix
elements, which determine the optical transition rates, assume the
following form in linear-in-$\beta$ approximation
\begin{eqnarray}\label{ev_k2}
&&|\langle + 1/2|\mathbf{e} \cdot \mathbf{v}|+3/2\rangle|^2 =
P^2/(2\hbar^2) \\
&& \hspace{1.2cm} \times [1 + 2\beta k_{y'}e_{x'}e_{y'} -
2\beta k_{x'}(e_{x'}^2-e_{y'}^2)] \:, \nonumber \\
&&|\langle -1/2|\mathbf{e} \cdot \mathbf{v}|-3/2\rangle|^2 =
P^2/(2\hbar^2) \nonumber \\
&& \hspace{1.2cm} \times [1 - 2\beta k_{y'}e_{x'}e_{y'} + 2\beta
k_{x'}(e_{x'}^2-e_{y'}^2)] \:. \nonumber
\end{eqnarray}
It follows from Eq.~(\ref{ev_k2}) that, for a fixed light
polarization, the spin-up and spin-down electrons are
predominantly photoexcited in opposite points in the $\mathbf{k}$
space. This is illustrated in Fig.~1(b) for the light polarized
along the $y'$ axis, where electrons with the spin $+1/2$ are
generated at a higher rate into states with positive values of
$k_{x'}$ whereas electrons with the spin $-1/2$ are mainly
generated into states with $k_{x'} < 0$. The difference in rates
is shown by vertical lines of different thicknesses. We note that
here the spin-orbit splitting of the subbands $hh1$ and $e1$ is
unimportant and, therefore, not shown in Fig.~1(b) for simplicity.
The spin-dependent asymmetry of optical excitation leads also to
the pure spin current. Calculation shows that, in (110)-grown QWs,
components of the spin photocurrent caused by ${\bf k}$-linear
terms in the matrix elements of optical transitions have the form
\begin{equation}\label{Jcv110b}
J_{x'}^{z'} = \beta (e_{y'}^2-e_{x'}^2) \frac{\tau_e
\varepsilon}{\hbar} \frac{\eta_{cv}}{\hbar\omega} I \:, \;\;
J_{y'}^{z'} = \beta e_{x'} e_{y'} \frac{\tau_e \varepsilon}{\hbar}
\frac{\eta_{cv}}{\hbar\omega} I \:,
\end{equation}
where $\varepsilon=(\hbar\omega-E_g^{QW})\mu/m_e$ is the kinetic
energy of photoexcited electrons. In contrast to Eq.~(1), this
contribution depends on the polarization plane of the incident
light and vanishes for unpolarized light. From comparison
of~Eqs.~(\ref{Jcv110a}) and~(\ref{Jcv110b}) one can see that,
depending on the value of $\hbar \omega - E^{QW}_g$, the two
contributions to $J_{x'}^{z'}$ can be comparable or one of them
can dominate over the other. We also note that the both spin
current contributions are caused by bulk inversion asymmetry and
do not vanish in symmetrically-grown (110)-QWs.

In (001)-grown QWs, the absorption of linearly polarized or
unpolarized light results in a in-plane flow of electron spins,
see Eq.~(\ref{J_phenC2v}). In contrast to the low-symmetry QWs
considered above, in the (001)-QW structures the
linear-in-$\mathbf{k}$ terms in the matrix elements of optical
transitions from the heavy-hole subband vanish at the normal
incidence. Since, in addition, the $\mathbf{k}$-linear spin
splitting of the heave-hole subband is suppressed in (001)-grown
structures~\cite{ivchbook,Winkler03}, we conclude that the spin
photocurrents are entirely related to spin-orbit splitting of the
conduction subband. Assuming the parabolic spin-independent
dispersion in the $hh1$ subband and taking into account the
spin-dependent contribution
\begin{equation}\label{Hso_e1}
H_{\mathrm{so}}^{(e1)} = \sum_{\alpha\beta} \gamma_{\alpha
\beta}^{(e1)} \, \sigma_{\alpha} k_{\beta}
\end{equation}
to the electron effective Hamiltonian in the subband $e1$, the
components of pure spin current generated in the subband $e1$ are
derived to be
\begin{equation} \label{001spin}
J^{\alpha}_{\beta} = \gamma_{\alpha\beta}^{(e1)} \frac{\tau_e\,
\mu}{2 \hbar m_h} \frac{\eta_{cv}}{\hbar \omega} I \:.
\end{equation}
For the interband transitions from the light-hole subband or the
spin-split band $\Gamma_7$, both the split-subband-related and the
matrix-element-related mechanisms lead to polarization-dependent
pure spin photocurrents. The analysis shows that, in the geometry
of normal incidence, the optical excitation from the light-hole
subband in (001)-grown QWs leads to the spin current described by
Eq.~(\ref{J_phenC2v}) where the phenomenological coefficients
satisfy the relations $L_4^B=L_2^B-L_1^B$ and $L_1^S=L_3^S-L_2^S$.
If the spin photocurrent is solely caused by $\mathbf{k}$-linear
terms in the matrix elements of optical transitions then, in
addition, $L_1^B=0$ and $L_2^S=L_3^S=0$. In the opposite case,
when the spin current is mainly contributed by the
split-subband-related mechanism, the coefficients are
interconnected by $L_1^B=\pm L_2^B$ and $L_2^S=\pm L_3^S$ with the
sign ``$+$'' or ``$-$'' depending on, respectively, whether the
spin splitting of the subbands $e1$ or $lh1$
predominates,~Ref.~\cite{Tarasenko05}.

The injection of pure spin currents in (110)-oriented GaAs QWs at
room temperature by one-photon absorption of a linearly polarized
optical pulse was demonstrated by Zhao et al.~\cite{Zhao05}.
Spatially resolved pump-probe technique was used which enabled the
authors to obtain signatures of the pure spin currents by
measuring the resulting spin separations of $1 \div 4$~nm. The
pump pulse excited electrons from the valence to the conduction
band with an excess energy of $\sim 148$~meV. The probe was tuned
near the band edge. It was observed that the spin current
resulting in separation of the spin density $S_{z'}$ along the
$[1\bar{1}0]$ axis reversed its direction when the polarization of
the pump pulse was switched from $\mathbf{e}\|x'$ to
$\mathbf{e}\|y'$. This indicates that, for the photon energy used
in the experiment, the polarization-dependent contribution
dominates over the polarization-independent term.

\section{Intersubband transitions in $N$-doped QWs}\label{Intersubband}

The intersubband light absorption in $n$-doped QW structures is a
resonant process which becomes possible if the photon energy
$\hbar\omega$ is tuned to the intersubband energy separation. In
the simple one-band model, direct optical transitions between the
electron subbands $e1$ and $e2$ conserve spin and are induced only
by radiation with a nonzero normal component $e_{\perp}$ of the
polarization vector. If the spin degeneracy of the quantum
subbands is lifted, such spin-conserving optical transitions give
rise to a pure spin current~\cite{TarIvch05,Sherman05}. This
mechanism is illustrated in Fig.~2, where the intersubband
transitions $(e1,+1/2)\rightarrow(e2,+1/2)$ and
$(e1,-1/2)\rightarrow(e2,-1/2)$ are shown by vertical solid lines.
\begin{figure}[b]
\leavevmode \epsfxsize=0.95\linewidth
\centering{\epsfbox{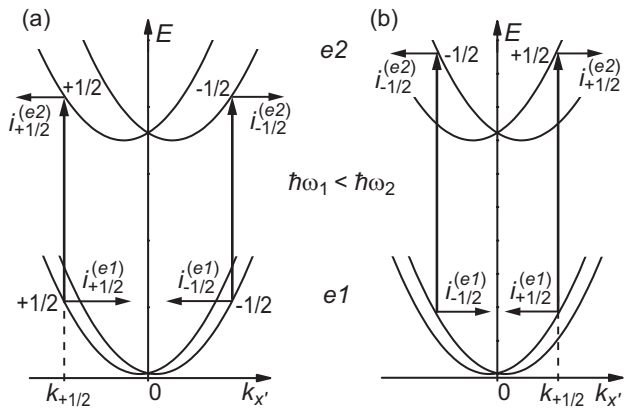}} \caption{Microscopic mechanism
of the pure spin photocurrent induced by intersubband excitation
with linearly polarized light due to $\mathbf{k}$-linear splitting
of the subbands. Panels (a) and (b) demonstrate the spin current
reversal with increasing the light frequency.}
\end{figure}
Due to $\mathbf{k}$-linear spin splitting of the subbands together
with the energy and quasi-momentum conservation, the optical
transitions induced by light of a fixed frequency occur only at
certain values of $k_{x'}$, denoted by $k_{+1/2}$ and $k_{-1/2}$
for the spin states $\pm1/2$, respectively, where the photon
energy $\hbar\omega$ matches the energy spacing between the
subbands. As is evident from Fig.~2(a), these $k_{x'}$-points are
of opposite signs for transitions from the spin branches $\pm
1/2$. Similarly to the interband light absorption considered in
Sect.~\ref{Interband}, such spin-dependent asymmetry of
photoexcitation gives rise to pure spin currents in both $e1$ and
$e2$ subbands.

An interesting feature of the pure spin photocurrent caused by
$\mathbf{k}$-linear splitting of the subbands is its spectral
response. Figures~2(a) and~2(b) show what happens if the photon
energy $\hbar\omega$ crosses the resonance varying from
$\hbar\omega < E_{21}$ to $\hbar\omega > E_{21}$, where $E_{21}$
is the energy separation between the subbands at $\mathbf{k}=0$.
For the photon energy below $E_{21}$ [see Fig.~2(a)], the optical
transitions $(e1,+1/2) \rightarrow (e2,+1/2)$ occur at negative
values of $k_{x'}$ leading to a flow of spin-up electrons in the
subband $e1$ in the $x'$ direction. With increasing the light
frequency, the point of optical transitions $k_{x'}=k_{+1/2}$ at
which the energy and quasi-momentum conservation laws are met
moves toward positive values of $k_{x'}$ [see Fig.~2(b)]. This
results in an inversion of the spin current.

The explicit spectral dependence of the spin photocurrent in an
ideal QW drastically depends on the fine structure of the energy
spectrum. In real QW structures, the spectral width of the
intersubband resonance is substantially broadened. Allowance for
the broadening can be made assuming, e.g., that the energy
separation $E_{21}$ between the subbands varies in the QW
plane~\cite{Ganichev03,Tarasenko07}. Then, to the first order in
the spin-orbit coupling, the spin current components in the
subbands are given by
\begin{equation}\label{J21_1a}
J^{\alpha(e1)}_{\beta} = \frac{\tau_{e1}\,e_{\perp}^2}{2\hbar}
\left(\gamma^{(e2)}_{\alpha\beta}-\gamma^{(e1)}_{\alpha\beta}\right)
\bar{E} \frac{\,d\,\eta_{21}(\hbar\omega)}{d\,\hbar\omega}
\frac{I}{\hbar\omega} \:,
\end{equation}
\begin{equation}\label{J21_2a}
\hspace{-1.6cm} J^{\alpha(e2)}_{\beta} =
\frac{\tau_{e2}\,e_{\perp}^2}{2\hbar}
\left(\gamma^{(e2)}_{\alpha\beta}-\gamma^{(e1)}_{\alpha\beta}\right)
\left[ \eta_{21}(\hbar\omega) \right.
\end{equation}
\vspace{-0.5cm}
\[ \hspace{3.5cm} \left. - \tau_{e2} \bar{E}\,
\frac{\,d\,\eta_{21}(\hbar\omega)}{d\,\hbar\omega} \right]
\frac{I}{\hbar\omega} \:,
\]
where $\tau_{e1}$ and $\tau_{e2}$ are the spin current relaxation
times in the subbands $e1$ and $e2$, respectively,
$\gamma^{(e1)}_{\alpha\beta}$ and $\gamma^{(e2)}_{\alpha\beta}$
are the constants of $\mathbf{k}$-linear spin-orbit coupling in
the subbands, see Eq.~(\ref{Hso_e1}), $\eta_{21}(\hbar\omega)$ is
the intersubband absorbance for radiation polarized along the QW
normal with the inhomogeneous broadening being taken into account,
and $\bar{E}$ is the mean value of the electron kinetic energy.
The energy $\bar{E}$ equals to $E_F/2$ for a two-dimensional
degenerate gas with the Fermi energy $E_F$ and $k_B T$ for a
non-degenerate gas at the temperature $T$.

The spin photocurrents~(\ref{J21_1a}) and~(\ref{J21_2a}) are
contributed by spin-conserving optical transitions and, therefore,
are proportional to the difference of subband splitting constants.
The spectral behavior of the pure spin currents in both subbands
repeats the derivative of the light absorption spectrum
$d\eta_{21}(\hbar\omega) / \hbar\omega$ provided the intersubband
absorption line is narrow enough. Close to the absorption maximum
the spin photocurrents reverse their directions with varying the
light frequency. We also note that the contribution
$J^{\alpha(e1)}_{\beta}$ can considerably exceed
$J^{\alpha(e2)}_{\beta}$ since the relaxation time in the excited
subband $\tau_{e2}$ may be quite short even at low temperatures
due to the effective channel of relaxation by emission of an
optical photon.

A contribution to the pure spin currents may also come from
linear-in-$\mathbf{k}$ spin-dependent terms in the matrix elements
of optical transitions. While in the one-band approximation the
intersubband absorption can only be induced by the $e_{\perp}$
component of the polarization vector, in a multi-band model
optical transitions between the electron subbands $e1$ and $e2$
are allowed for any polarization~\cite{Warburton96}. Moreover,
$\mathbf{k}$$\cdot$$\mathbf{p}$ admixture of the valence-band and
remote conduction-band states to the electron wave functions adds
both spin-dependent~\cite{Ivchenko04} and $\mathbf{k}$-linear
terms to the matrix elements of the optical transitions. Taking
into account these contributions, the $2\times2$ spin matrix
$M_{21}$ describing the intersubband transitions assumes the form
\[
M_{21} = M_{21}^{(0)} \left(e_{\perp} + i \sum_{\alpha\beta}
\lambda_{\alpha\beta} \, \sigma_{\alpha} e_{\beta} + i
\sum_{\alpha\beta} \lambda'_{\alpha\beta} \, k_{\alpha} e_{\beta}
\right.
\]
\vspace{-0.7cm}
\begin{equation}\label{M21}
\hspace{4cm} \left. + \sum_{\alpha\beta\gamma}
\lambda''_{\alpha\beta\gamma} \sigma_{\alpha} k_{\beta} e_{\gamma}
\right) \:,
\end{equation}
where $M_{21}^{(0)}$ is the matrix element calculated in the
one-band approximation for radiation polarized along the QW
normal. The tensor $\lambda$ is responsible for the intersubband
optical orientation of electron spins~\cite{Ivchenko04}, while
$\lambda'$ and $\lambda''$ describe the optical alignment of
electron momenta. Taking into account $\mathbf{k}$-linear terms in
Eq.~(\ref{M21}), we derive the contributions to pure spin currents
excited in the $e1$ and $e2$ subbands as
\begin{equation}\label{J21_1b}
J^{\alpha(e1)}_{\beta} = \left( \sum_{\gamma\delta}
\lambda_{\alpha\gamma} \lambda'_{\beta\delta} \,
e_{\gamma}e_{\delta} - \sum_{\gamma} \lambda''_{\alpha\beta\gamma}
\, e_{\perp}e_{\gamma} \right)
\end{equation}
\vspace{-0.3cm}
\[
\hspace{4cm} \times \frac{\tau_{e1} \bar{E}}{\hbar}\: \frac{I
\eta_{21}(\hbar\omega)}{\hbar \omega} \:,
\]
$J^{\alpha(e2)}_{\beta}=-(\tau_{e2}/\tau_{e1})J^{\alpha(e1)}_{\beta}$.
In contrast to Eq.~(\ref{J21_1a}), the spectral response of the
contribution~(\ref{J21_1b}) repeats the light absorption spectrum.

\section{Free-carrier absorption in $N$-doped QWs}\label{Intrasubband}

The light absorption by free carriers, or Drude-like absorption,
occurs in doped semiconductor structures when the photon energy
$\hbar\omega$ is smaller than the band gap as well as the energy
spacing between the subbands. Such an intrasubband excitation of
carriers with linearly polarized light also gives rise to a pure
spin current. However, in contrast to the direct transitions
considered in Sects.~\ref{Interband} and~\ref{Intersubband}, the
subband spin splitting leads to no essential contribution to the
spin current induced by intrasubband optical excitation. The more
important contribution comes from the spin-dependent asymmetry of
electron scattering~\cite{TarIvch05,Ganichev06}. Indeed, the
free-carrier absorption is always accompanied by electron
scattering from acoustic or optical phonons, static defects, etc.,
because of the need for energy and momentum conservation. In
systems with a spin-orbit interaction, processes involving change
of the particle wave vector are spin dependent. In particular, in
the QW structures the matrix element of electron scattering
$V_{\mathbf{k}'\mathbf{k}}$ contains, in addition to the main
contribution $V_0$, asymmetric spin-dependent
terms~\cite{Ivchenko04,Averkiev02}
\begin{equation}\label{V_sc}
V_{\mathbf{k}'\mathbf{k}} = V_0 + \sum_{\alpha\beta}
V_{\alpha\beta} \, \sigma_{\alpha}(k_{\beta}+k'_{\beta}) \:,
\end{equation}
where $\mathbf{k}$ and $\mathbf{k}'$ are the initial and the
scattered in-plane wave vectors, respectively. This leads in turn
to $\mathbf{k}$-linear spin-dependent contribution to the
scattering rate, which is determined by the matrix element
squared. Microscopically, such terms in the scattering rate
originate from structure and/or bulk inversion asymmetries similar
to $\mathbf{k}$-linear Rashba and Dresselhaus spin splitting of
the electron subbands.

Due to the spin-dependent asymmetry of scattering, electrons
photoexcited from the subband bottom are scattered in preferred
directions depending on their spin states. This is illustrated in
Fig.~3(a), where the free-carrier absorption is shown as a
combined two-stage process involving the electron-photon
interaction (solid vertical lines) and the electron scattering
(dashed horizontal lines). The scattering asymmetry is shown by
dashed lines of different thicknesses: electrons with the spin
$+1/2$ are preferably scattered into the states with $k_{x'}>0$,
while electrons with the spin $-1/2$ are predominantly scattered
into the states with $k_{x'}<0$. Obviously, such an asymmetry of
photoexcitation in the $\mathbf{k}$ space leads to a pure spin
current, where the spin-up and spin-down electrons counter flow
and the charge current vanishes.
\begin{figure}[b]
\leavevmode \epsfxsize=0.95\linewidth
\centering{\epsfbox{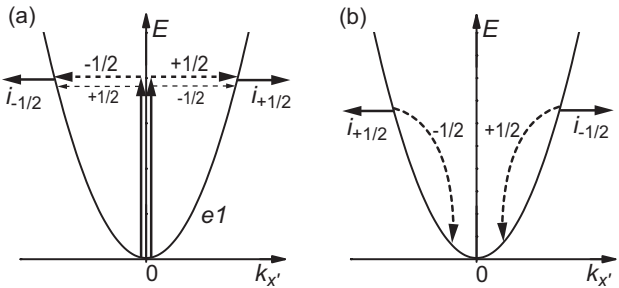}} \caption{Microscopic mechanisms
of the pure spin currents caused by spin-dependent scattering
processes at (a) intrasubband photoexcitation with linearly
polarized light and (b) energy relaxation of hot carriers.}
\end{figure}

In the perturbation theory approach, the indirect optical
transitions are treated as second-order virtual processes
involving intermediate states. To the first order in spin-orbit
interaction, the compound matrix element of the intrasubband
transitions accompanied by the electron scattering from
short-range potentials has the form~\cite{Tarasenko06}
\begin{equation}\label{M_e1}
M_{\mathbf{k}'\mathbf{k}} = \frac{eA}{c\,\omega m_e}\,
\mathbf{e}\cdot(\mathbf{k}'-\mathbf{k}) V_{\mathbf{k}'\mathbf{k}}
- 2\frac{eA}{c\hbar} \sum_{\alpha\beta}
V_{\alpha\beta}\,\sigma_{\alpha}e_{\beta} \:,
\end{equation}
where $e$ is the electron charge, $A$ is the vector potential of
the electromagnetic wave, and $c$ is the light velocity. The first
term on the right-hand side of Eq.~(\ref{M_e1}) describes
transitions $(e1,\mathbf{k})\rightarrow(e1,\mathbf{k}')$ with
intermediate states in the conduction subband $e1$, the second
term corresponds to the transitions via intermediate states in
other bands. We assume that the electron scattering is elastic and
consider the geometry of normal incidence of the light so that the
polarization vector $\mathbf{e}$ lies in the $(xy)$ plane. Then,
the polarization dependencies of spin current components are given
by
\begin{equation}\label{Je1}
J_{x}^{\alpha} = - \frac{\tau_e}{\hbar} \left( \frac{\langle V_0
V_{\alpha x} \rangle}{\langle V_0^2 \rangle}
\,\frac{e_x^2-e_y^2}{2} + \frac{\langle V_0 V_{\alpha y}
\rangle}{\langle V_0^2 \rangle} e_x e_y \right) I \eta_{e1} \:.
\end{equation}
Here, the angle brackets $\langle...\rangle$ stand for averaging
over the spacial distribution of scatterers and $\eta_{e1}$ for
the radiation absorbance in this spectral range. The components
$J_{y}^{\alpha}$ can be obtained from Eq.~(\ref{Je1}) by the
replacement $x \leftrightarrow y$.

Equation~(\ref{Je1}) shows that pure spin currents can be injected
in QWs by the elastic-scattering-assisted photoexcitation with
linearly polarized light but vanish for the normally-incident
unpolarized radiation, when
$\overline{e_x^2}=\overline{e_y^2}=1/2$, $\overline{e_xe_y}=0$.
The nonzero components of the spin current are determined by the
explicit form of the matrix element of scattering and the light
polarization plane. In QWs grown on (110)-oriented substrates, the
scattering rate contains the term proportional to $\langle V_0
V_{z'x'}\rangle \,\sigma_{z'}(k_{x'}+k'_{x'})$ giving rise to the
components $J_{x'}^{z'}\propto (e_{x'}^2-e_{y'}^2)$,
$J_{y'}^{z'}\propto e_{x'}e_{y'}$, which are in accordance with
the phenomenological equation~(\ref{J_phenCs}). In (001)-grown
structures, the nonzero coefficients are $\langle V_0
V_{xy}\rangle=-\langle V_0 V_{yx}\rangle$ and $\langle V_0
V_{xx}\rangle=-\langle V_0 V_{yy}\rangle$, and the
normally-incident radiation can excite fluxes of the in-plane spin
components only.

The pure spin current caused by the free-carrier absorption can be
converted into an electric current by polarizing electron spins,
e.g., by application of an external magnetic field, as was shown
by Ganichev et al.~\cite{Ganichev06}, see also~\cite{Belkov05}.
Indeed, in the case of intrasubband absorption, the fluxes of the
spin-up and spin-down electrons, $\mathbf{i}_{+1/2}$ and
$\mathbf{i}_{-1/2}$, are proportional to the electron densities in
the spin subbands, $n_{+1/2}$ and $n_{-1/2}$, respectively. In a
spin-polarized system, where $n_{+1/2}\neq n_{-1/2}$, the fluxes
$\mathbf{i}_{+1/2}$ and $\mathbf{i}_{-1/2}$ do no longer
compensate each other yielding a net electric current
\begin{equation}\label{j_electric}
j_{\beta} = 4 e \sum_{\alpha} S_{\alpha} J_{\beta}^{\alpha} \:,
\end{equation}
where $\mathbf{S}$ is the average electron spin with
$|\mathbf{S}|=(1/2)|n_{+1/2}-n_{-1/2}|/(n_{+1/2}+n_{-1/2})$.

\section{Pure spin currents caused by electron gas
heating}\label{Heating}

In addition to the free-carrier absorption, the spin-dependent
asymmetry of the electron scattering by phonons gives rise to a
pure spin current if the electron gas is simply driven out of
thermal equilibrium with the crystal lattice~(see
Refs.~\cite{Belkov05,Ganichev06,Ganichev07}). In such a
relaxational mechanism, the spin current is generated in the
process of energy relaxation of electrons no matter how the
thermal equilibrium between the electron and phonon subsystems was
initially disturbed.

The relaxational mechanism of the spin current generation is
illustrated in Fig.~3(b), where the processes of energy relaxation
of hot electrons by emitting phonons are shown by dashed curves.
Due to the spin-dependent asymmetry of the electron-phonon
interaction, electrons with the spin $+1/2$ relax faster from the
high-energy states with positive $k_{x'}$, while electrons with
the spin $-1/2$ predominantly vacate the high-energy states with
negative $k_{x'}$. This leads to an asymmetrical distribution,
where the spin-up carriers occupy mainly the left-hand branch of
the dispersion curve (carriers with the opposite spin orientation
have gone to the subband bottom), while the spin-down carriers
occupy mainly the right-hand branch. Such a spin-dependent
imbalance of electrons between positive and negative $k_{x'}$
yields a pure spin current.

We consider the energy relaxation of electrons confined in a QW by
bulk acoustic phonons. Taking into account $\mathbf{k}$-linear
contributions to the electron-phonon interaction, the matrix
element of the electron scattering by phonons can be modeled by
\begin{equation}\label{V_mod}
V_{\mathbf{k}'\mathbf{k}}(\mathbf{q}) = \mathcal{V}_0(q_{\perp}) +
\sum_{\alpha\beta} \mathcal{V}_{\alpha\beta}(q_{\perp})
\,\sigma_{\alpha}(k_{\beta}+k'_{\beta}) \:,
\end{equation}
where $\mathcal{V}_0(q_{\perp})$ and
$\mathcal{V}_{\alpha\beta}(q_{\perp})$ are functions of
$q_{\perp}$ with their forms dependent of the QW design, and
$\mathbf{q}=\pm(\mathbf{k}-\mathbf{k}',q_{\perp})$ is the
three-dimensional wave vector of the phonon involved. We assume
that both electrons and phonons obey the Boltzmann statistics, but
the electron temperature $T_e$ differs from the lattice
temperature $T_0$. Then, the rates of phonon emission and
absorption become nonequal leading to a spin current
\begin{equation}\label{J_heat1}
J_{\beta}^{\alpha} =  - \frac{N_e}{2} \frac{\tau_e}{\tau_{ph}}
\frac{\hbar \,c_s^2}{k_B T_0} \frac{T_e - T_0}{T_e} \times
\end{equation}
\vspace{-1cm}
\[
\times \frac{\int\limits_{-\infty}^{+\infty}
\mathrm{Re}[\mathcal{V}_0^*(q_{\perp})
\mathcal{V}_{\alpha\beta}(q_{\perp})] |q_{\perp}|
dq_{\perp}}{\int\limits_{-\infty}^{+\infty}
|\mathcal{V}_0(q_{\perp})|^2 /|q_{\perp}| \, dq_{\perp}} \:,
\]
where $N_e$ is the carrier density, $\tau_{ph}$ is the momentum
relaxation time governed by the electron-phonon interaction, and
$c_s$ is the sound velocity in the crystal.

As a more detailed example, we consider the (110)-grown QWs. In
this case the dominant spin-dependent contribution to the
Hamiltonian of electron-phonon interaction in the
deformation-potential model is proportional to
$\sigma_{z'}(k_{x'}+k'_{x'})$, and the corresponding Hamiltonian
has the form~\cite{Tarasenko08}
\begin{equation}\label{V_elphon}
H_{\mathrm{el-phon}}(\mathbf{k}',\mathbf{k}) = \Xi_c \sum_{\alpha}
u_{\alpha\alpha} + \xi \, \Xi_{cv} \,u_{z'z'} \sigma_{z'} (k_{x'}
+ k'_{x'})/2 \:.
\end{equation}
Here, $\Xi_c$ and $\Xi_{cv}$ are the intraband and interband
constants of the deformation potential, $u_{\alpha\beta}$ are the
phonon-induced strain tensor components, $\xi=P
\Delta_{\mathrm{so}}/[3E_g(E_g+\Delta_{\mathrm{so}})]$, and
$\Delta_{\mathrm{so}}$ is the spin-orbit splitting of the valence
band. The interband constant $\Xi_{cv}$ originates from the lack
of an inversion center in zinc-blende-type crystals and vanishes
in centrosymmetric semiconductors~\cite{optor}. Assuming that
electrons are confined in a rectangular quantum well of the width
$a$, we derive for the spin current
\begin{equation}\label{J_heat2}
J_{x'}^{z'} = - \frac{\pi^2 \xi}{3 a^2} \frac{\tau_e}{\tau_{ph}}
\frac{\hbar \,c_s^2}{k_B T_0} \frac{\Xi_{cv}}{\Xi_c} \,
\frac{T_e-T_0}{T_e} N_e \:.
\end{equation}
Equation~(\ref{J_heat2}) shows that the spin current component
$J_{x'}^{z'}$ strongly depends on the QW width.

\section{Pure valley-orbit currents}\label{Valley}
\vspace{-0.1cm}
In addition to the spin, free carriers in solid states can be
characterized by another internal property, e.g., by a well number
in multiple QW structures or a valley index $\nu$ in many-valley
semiconductors. In the latter case, one can consider pure
orbit-valley currents, where partial electron fluxes in valleys
$\mathbf{i}_{\nu}$ are nonzero but the net electric current $e
\sum_{\nu} \mathbf{i}_{\nu}$ vanishes~\cite{TarIvch05}. Here, the
role of spin-up and spin-down states is replaced by the valley
index: there is no net charge current, but the electrons in
different valleys travel in different directions.

To elaborate the concept of pure orbit-valley currents, we
consider silicon-based quantum wells grown on a (111)-oriented
surface. In Si QWs, the conduction-band subbands are formed by six
equivalent valleys $X$, $X'$, $Y$, $Y'$, $Z$, and $Z'$ located at
the $\Delta$ points of the Brillouin zone of the bulk crystal. All
the valleys retain their equivalence in (111)-grown structures
because the angles between the growth direction and the valley
principle axes are the same. Figure~4 sketches the valley
positions and orientations in the two-dimensional $\mathbf{k}$
space in the QW plane. In asymmetrical (111)-grown QWs, each
valley has the $C_s$ point-group symmetry allowing for the
generation of a partial in-plane flux $\mathbf{i}_{\nu}$ at normal
incident of the light. Under excitation with unpolarized light,
the fluxes $\mathbf{i}_{\nu}$ are directed along the in-plane
projections of the valley principle axes [see Fig.~4(a)]. Since
the structure is invariant with respect to the rotation by
$120^\circ$ along the growth direction, the total charge current
$e\sum_{\nu} \mathbf{i}_{\nu}$ vanishes. Thus, such an electron
distribution can be referred to as an optically injected pure
valley-orbit current.
\begin{figure}[t]
\leavevmode \epsfxsize=0.95\linewidth
\centering{\epsfbox{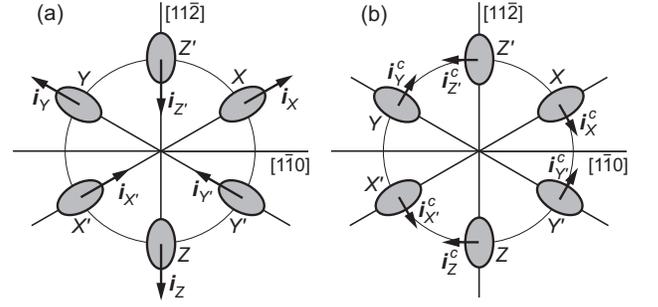}} \caption{(a) Angular
distribution of electron fluxes in valleys under excitation with
unpolarized light. (b) Angular distribution of helicity-dependent
fluxes in valleys.}
\end{figure}

In addition to the polarization-independent photocurrent, the
excitation of a Si (111)-grown QW with circularly polarized light
at normal incidence results, in each valley, in a flux component
$\mathbf{i}_{\nu}^c$ which reverses its direction upon switching
the light polarization from right-handed to left-handed circular
polarization. Such helicity-dependent components
$\mathbf{i}_{\nu}^c$ flow perpendicularly to the valley principle
axes [see Fig.~4(b)] and also contribute to the pure valley-orbit
current. We note that the absence of a total photocurrent under
illumination with unpolarized or circularly polarized light is
related to the overall $C_{3v}$ symmetry of the QWs which,
however, allows for a net electric current induced by linearly
polarized light. In this particular case, the partial fluxes in
valleys become nonequal and do not compensate each other.

Pure valley-orbit currents can also be optically injected in bulk
multi-valley noncentrosymmetrical crystals such as AlAs, AlSb,
GaP, etc. In these compounds, the conduction-band minima are
located in the $X$ points at the Brillouin-zone edge. Each of
three equivalent valleys $\nu=X,Y,Z$ has the $D_{2d}$ symmetry
allowing for the helicity-dependent electron flux
$\mathbf{i}_{\nu}$
\begin{eqnarray}\label{valley-current}
&&\mathbf{i}_X = {\cal P} I (0, \ae_y, - \ae_z)\: ,\\
&&\mathbf{i}_Y = {\cal P} I(- \ae_x, 0, \ae_z)\: , \nonumber\\
&&\mathbf{i}_Z = {\cal P} I(\ae_x, -\ae_y, 0)\: , \nonumber
\end{eqnarray}
where $\ae_{\alpha}$ $(\alpha = x, y, z)$ are components of the
vector ${\rm i} [\mathbf{e} \times \mathbf{e}^*] =
P_{circ}\,\mathbf{q}/q$, $\mathbf{q}$ is the light wave vector,
and $P_{circ}$ is the light helicity ranging from $-1$ to $+1$. In
accordance with the overall $T_d$ point group of the
zinc-blende-type crystals, the total current vanishes for the
homogenous illumination with circularly polarized light. In an
external magnetic field, each contribution $\mathbf{i}_{\nu}$
varies due to the Lorentz force acting upon electrons. This action
is, however, different for different valleys due to the energy
spectrum anisotropy in valleys. As a result, the magnetic field
causes an imbalance of the valley-orbit current giving rise to a
nonzero net electric current
\[
\mathbf{j} \propto (\ae_y B_z + \ae_z B_y , \ae_z B_x + \ae_x B_z
, \ae_x B_y + \ae_y B_x)\:.
\]
For the particular geometry $\mathbf{q} \| [111]$ and $\mathbf{B}
\| [010]$, the magnetic field induced photocurrent $\mathbf{j}$
appears in the $[101]$ direction.

\section{Conclusion}

We have shown that pure spin currents of free carriers can readily
be created in semiconductor structures by optical excitation with
linearly polarized or even unpolarized light. The pure spin
currents lead to spacial separation of the spin-up and spin-down
particles and accumulation of the opposite spins at the opposite
edges of the sample. We have presented the microscopic theory of
pure spin photocurrents for all main types of optical transitions
ranging from the fundamental interband to the free-carrier
absorption. In the present paper we have focused on the spin
photocurrents contributed by charge carriers. In addition, spin
fluxes (or, in general, angular-momentum fluxes) can also be
formed by neutral particles or excitations lacking electric charge
such as photons~\cite{Barnett02,Alexeyev08}, excitons or exciton
polaritons~\cite{Kavokin05,Leyder07}, and even phonons and
magnons. The study of spin currents is naturally inscribed in the
physics of spin-related phenomena and opens up new opportunities
for the realization of novel device concepts.

\textbf{Acknowledgement} This work was supported by the Russian
Foundation for Basic Research, programs of the Russian Academy of
Sciences, and the Council of the President of the Russian
Federation for Support of Young Scientists.


\begin{thebibliography}{99}
\bibitem{sc_drag} Note that possible contributions to the spin
photocurrent caused by transfer of the photon linear momenta is
ignored in Eq.~(\ref{J_phen}).
\end{thebibliography}
\end{document}